\documentclass[reprint,superscriptaddress,amsmath,amssymb,aps,prb]{revtex4-1}

\usepackage[version=3]{mhchem} 
\usepackage{graphicx}
\usepackage{hyperref}
\usepackage{physics}
\usepackage{braket}
\usepackage{siunitx}

\newcommand{\code}[1]{\texttt{\detokenize{#1}}}
\newcommand{\JSC}[0]{$J_\text{SC}$}
\newcommand{\VOC}[0]{$V_\text{OC}$}

\begin{document}

\title{Upper limit to the photovoltaic efficiency of imperfect crystals: the case of kesterite solar cells}

\author{Sunghyun Kim}
\affiliation{Department of Materials, Imperial College London, Exhibition Road, London SW7 2AZ, UK} 

\author{Jos\'e A. M\'arquez}
\affiliation{Helmholtz-Zentrum Berlin f\"ur Materialien und Energie GmbH, Department Structure and Dynamics of Energy Materials, Hahn-Meitner-Platz 1, D-14109 Berlin, Germany}

\author{Thomas Unold}
\affiliation{Helmholtz-Zentrum Berlin f\"ur Materialien und Energie GmbH, Department Structure and Dynamics of Energy Materials, Hahn-Meitner-Platz 1, D-14109 Berlin, Germany}
\email{unold@helmholtz-berlin.de}

\author{Aron Walsh}
\affiliation{Department of Materials, Imperial College London, Exhibition Road, London SW7 2AZ, UK} 
\affiliation{Department of Materials Science and Engineering, Yonsei University, Seoul 03722, Korea}
\email{a.walsh@imperial.ac.uk}
\date{\today}

\begin{abstract}
The Shockley-Queisser (SQ) limit provides a convenient metric for predicting light-to-electricity conversion efficiency of a solar cell based on the band gap of the light-absorbing layer. In reality, few materials approach this radiative limit. We develop a formalism and computational method to predict the maximum photovoltaic efficiency of imperfect crystals from first principles. Our scheme includes equilibrium populations of native defects, their carrier-capture coefficients, and the associated recombination rates. When applied to kesterite solar cells, we reveal an intrinsic limit of 20\% for \ce{Cu2ZnSnSe4}, which falls far below the SQ limit of 32\%. The effects of atomic substitution and extrinsic doping are studied, leading to pathways for an enhanced efficiency of 31\%. This approach can be applied to support targeted-materials selection for future solar-energy technologies. 
\end{abstract}

\maketitle
Sunlight is the most abundant source of sustainable energy.
Similar to the Carnot efficiency of heat engines,
the maximum efficiency for photovoltaic energy conversion is determined by thermodynamics
and can be as high as 86\% owing to the high temperature of the sun.\cite{Landsberg:1980hy, Marti:1996bh}
However, in practical solar cells with single \textit{p-n} semiconductor junctions,
large irreversible energy loss occurs mainly through hot-carrier cooling and low light absorption below the band gap.\cite{sq}

The Shockley-Queisser (SQ) limit describes the theoretical sunlight-to-electricity conversion efficiency of a single-junction solar cell.\cite{sq} 
The SQ limit (33.7\% under AM1.5g illumination) and its variations, including spectroscopic limited maximum efficiency (SLME),\cite{slme}
determine the maximum efficiency of a solar cell based on the principle of \emph{detailed balance} between the absorption and emission of light.
The amount of photons absorbed determines the short-circuit current density $J_\text{SC}$, 
and, hot-carrier cooling and radiative recombination limit the maximum carrier concentration and hence the open-circuit voltage $V_\text{OC}$.

In the SQ limit, the predicted efficiency is a function of the semiconductor band gap, 
which is a trade-off between light absorption (current generation) and energy loss due to hot-carrier cooling.
This analysis secured the band gap as a primary descriptor when searching for 
new photovoltaic compounds, often within a 1--1.5 eV target window.
Unfortunately, few materials approach the SQ limit. 
Less than 10 classes of materials have achieved 
conversion efficiency greater than 20\%.\cite{pvtables}
Most emerging technologies struggle to break the 10\% efficiency threshold. 

Kesterites are a class of quaternary materials studied for thin-film photovoltaic applications.
Although a lot of progress has been made during the past few decades, 
the certified champion efficiency of 12.6\%\cite{Son:2019hm} has been increased by less than 0.1\% since 2013.\cite{Wang:2013gs}
The main bottleneck is the low open-circuit voltage, which is far below the SQ limit.\cite{Wallace:2017hga}
Many routes to engineer compositions and architectures have been considered, 
but it is not clear which process dominates.\cite{Grenet:2018ia}
One of the biggest questions in the field is if there is an intrinsic problem with kesterite semiconductors that prevent them approaching the radiative limit.\cite{siebentritt2013kesterite,bourdais2016cu,schorr2019point}

The discrepancy between the SQ limit and efficiencies of real solar cells results from the extra irreversible processes such as electron-hole nonradiative recombination.
While Shockley and Queisser studied the effect of the nonradiative recombination,
it has been treated as a parameter of radiative efficiency and often a radiative efficiency of 100\% is assumed,
which is unrealistic for real materials.

The rate of nonradiative recombination mediated by traps can be described by Shockley-Read-Hall statistics.\cite{Shockley:1952it, Hall:1952iz}
The steady-state recombination rate is determined by the detailed balance where the net electron-capture rate is equal to the net hole capture rate. 
A microscopic theory of carrier capture was proposed by Henry and Lang in 1977.\cite{Henry:1977bfa} 
The thermal vibration of the defect, together with the electron-phonon coupling, causes charge transfer from a delocalised free carrier to a localised defect state.
Thus the carrier capture coefficient heavily depends on the electron and phonon wave functions associated with a defect,
which are difficult to probe experimentally.
Instead, the microscopic processes in materials, including nonradiative carrier capture, 
have been inferred from macroscopic responses such as a capacitance transient.\cite{Henry:1977bfa}
%
%
Macroscopic properties of solar cells (e.g. open-circuit voltage and device efficiency) and 
microscopic processes in the material (e.g. carrier capture coefficient) are rarely connected.
Therefore, although theories of solar cells are well known, 
the theoretical approaches have failed to provide \textit{a priori} predictions of photovoltaic efficiencies of real materials.

Each material has a fundamental limit of radiative efficiency 
because the material contains a certain amount of native defects.
Their concentrations in thermal equilibrium are intrinsic properties of the materials, and
the resulting `soup' of defects determines the maximum radiative efficiency. 
Recently, first-principles methods based on density functional theory (DFT) have been developed to calculate the nonradiative carrier capture,\cite{Shi:2012iv,Alkauskas:2014kk,Kim:2019hp}
which opens up the possibility for studying the theoretical upper-bound of photovolataic efficiency of a real material
limited by both the radiative and the nonradiative recombination.

\begin{figure}[t]
    \centering
    \includegraphics[width=\columnwidth]{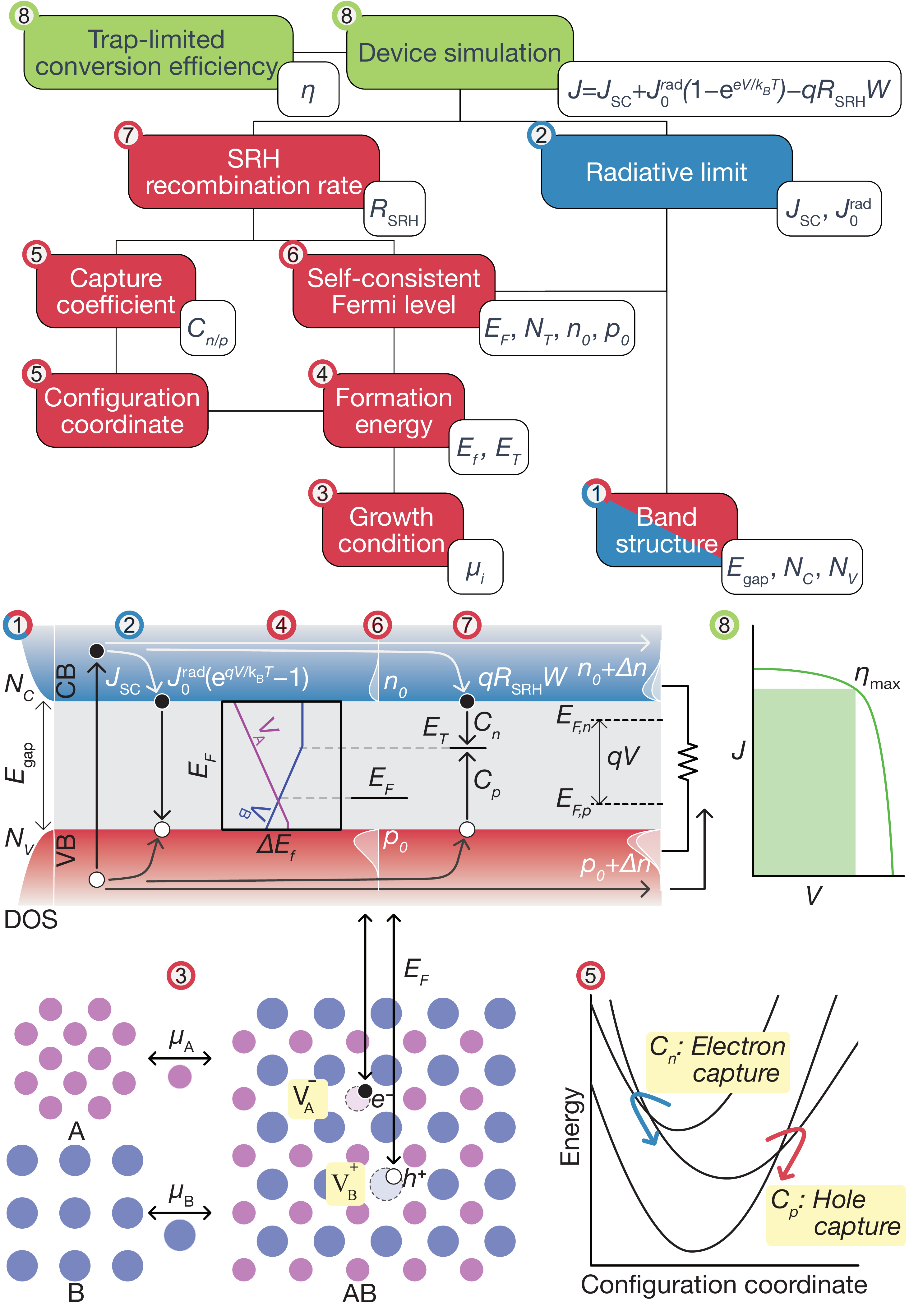}
    \caption{
    \textbf{Diagram for the calculation of trap-limited conversion efficiency.}
    The dependent calculations are connected by lines (upper panel).
    For each numbered step, the calculated quantities are appended. 
    The red and blue boxes represent calculations for radiative and nonradiative electron-hole recombination, respectively.
    The combined device simulations are marked in green.
    The corresponding physical processes are drawn in the electronic and the atomic structures (lower panel).
    }
    \label{fig:method}
\end{figure}

In this work, we propose a first-principles method of the \emph{trap-limited conversion efficiency} (TLC)
to calculate the upper-limit of photovolatic efficiency of a material containing the number of native defects in thermal equilibrium.
To take into account both radiative and nonradiative processes,
we perform a series of calculations for kesterites.
The absorption and the emission of light are calculated in the framework of Shockley and Queisser.
To obtain the nonradiative recombination rate, we calculate the carrier capture coefficients and equilibrium concentrations of native defects.
The workflow for our method is shown in Fig. \ref{fig:method}.
We conclude that kesterite solar cells suffer from significant nonradiative recombination and are unable to reach the SQ limit even under optimal growth conditions. 
Strategies to overcome such rapid recombination rates are suggested.

\section{Theory}
\subsection{Radiative Recombination}

The short-circuit current $J_\mathrm{SC}$ of a solar cell whose absorber thickness is $W$
is given by the absorbed photon flux multiplied by an elementary charge $\mathrm{q}$:
\begin{equation}\label{JSC}
    J_\mathrm{SC}\left( W \right) = \mathrm{q} \int_{0}^{\infty} %
    a \left( E; W \right) 
    \Phi_\mathrm{sun}\left( E \right) \, \dd E, 
\end{equation}
where $\Phi_\mathrm{sun}(E)$ and $a(E; W)$ are the solar spectrum and the absorptivity at a photon energy $E$, respectively. 
Following the SQ limit, we assume that an absorbed photon generates one electron-hole pair.

The radiative recombination rate for the solar cell at temperature $T$ is given by 
\begin{equation}
\begin{split}
    R_\mathrm{rad}(V) = & \frac{2\pi}{c^2 h^3} \int_{ 0 }^{\infty}
    a \left( E; W \right) \left[ \mathrm{e}^{\flatfrac{E - \mathrm{q}V}{k_B T}}-1 \right]^{-1} 
    E^2 \dd{E} \\ \approx &
    \frac{2\pi}{c^2 h^3} \mathrm{e}^{\frac{\mathrm{q}V}{k_{B} T}} \int_{ 0 }^{\infty}
    a \left( E;W \right) \left[ \mathrm{e}^{\flatfrac{E}{k_B T}}-1 \right] ^{-1}
     E^2 \dd{E} \\ = &
     R_\mathrm{rad}(0) \mathrm{e}^{\frac{\mathrm{q}V}{k_{B} T}} ,
\end{split}
\end{equation}
where $V$ is a bias voltage serving a chemical potential of the electron-hole pair.
At the short-circuit condition, the solar cell and ambient are in equilibrium:
the radiative recombination rate $R_\mathrm{rad}(0)$ is equal to the absorption rate from the ambient irradiation.
The net current density $J^\text{rad}$ limited by the radiative recombination is given by
\begin{equation}\label{eq:SQJV}
    J(V; W) = J_\mathrm{SC}(W) + J_{0}^\mathrm{rad}(W)(1 - \mathrm{e}^{\frac{\mathrm{q}V}{k_B T}}),
\end{equation}
where the saturation current $J_{0}^\text{rad} = \mathrm{q} R_\mathrm{rad}(0)$.

In the SQ limit, an absorptivity is assumed to be a step function being 1 above the band gap $E_g$ and 0 otherwise,
while a real material has a finite absorptivity with a tail near the band gap $E_g$, which depends on the sample thickness.
Rau \textit{et al.} \cite{Rau:2017ea} defined a \textit{photovoltaic} band gap using the absorption edge spectrum
and found that, in inorganic solar cells, the effect of the finite absorption tail on the open-circuit voltage loss is small.\cite{Rau:2017ea}
However, the band tail due to the disorder can cause serious reduction in \VOC~.

\subsection{Nonradiative Recombination}
A material in thermal equilibrium will contain a population of native defects. 
Defect processes are unavoidable and define the upper limit of performance of optoelectronic devices.
The nonradiative recombination at charge carriers via defects is often a dominant source of degradation of solar cells and should be carefully controlled.\cite{Park:2018et}

Based on the principle of detailed balance \cite{Shockley:1952it, Hall:1952iz}, 
the steady-state recombination rate $R_\mathrm{SRH}$ via a defect with electron-capture coefficient $C_{n}$ and hole-capture cross coefficient $C_{p}$ is given by 
\begin{equation}
\label{eq:SRH}
R_\mathrm{SRH}=\frac{np-n^2_{i}}{\tau_{p}(n+n_{t})+\tau_{n}(p+p_{t})},
\end{equation}
where 
\begin{equation}
\begin{split}
\tau^{-1}_{n}=N_{T}C_{n} = N_{T} \sigma_{n} v_{\mathrm{th},n},  \\
\tau^{-1}_{p}=N_{T}C_{p} = N_{T} \sigma_{p} v_{\mathrm{th},p}.
\end{split}
\end{equation}
Here, $n$, $p$, and $N_{T}$ denote concentrations of electrons, holes, and defects, respectively.
$n_i$ is an intrinsic carrier concentration ($n_i^2 = n_0 p_0$, where $n_0$ and $p_0$ are intrinsic electron and hole concentrations).
$n_{t}$ and $p_{t}$ represent the densities of electrons and holes, respectively, when the Fermi level is located at the trap level $E_{T}$.
The capture cross section ($\sigma_{n}$ for electron and $\sigma_{p}$ for hole) is commonly used in experimental studies,
and can be calculated taking the thermal velocities of electron $v_{\mathrm{th},n}$ and hole  $v_{\mathrm{th},p}$ to be \SI{E7}{\cm\per\s}.

For doped semiconductors, minority carrier lifetime often determines the rate of the total recombination process.
For example, in a \textit{p}-type semiconductor
where the acceptor concentration, $p_{0}$, is much higher than the 
photoexcited carrier density, the $R_\mathrm{SRH}$ due to a deep defect
is proportional to the (photoexcited) excess carrier density $\Delta n$:\cite{Nelson:2003is}
\begin{equation}\label{eq:SRHp}
R_\text{SRH} \approx \frac{\Delta n}{\tau_{n}} =  \Delta n N_{T} C_{n}.
\end{equation}
In case of a material containing many types of recombination centers, 
the total recombination rate $R_\text{SRH}$ is the sum over all independent centers.

The calculation of $R_\mathrm{SRH}$ requires three properties of a defect (concentration $N_{T}$, defect level $E_{T}$, and capture coefficient $C_{n/p}$)
in addition to the carrier concentrations $n$ and $p$, as well as the intrinsic doping density $n_{0}$ or $p_{0}$ in the bulk host,
as explained in the following subsections.

\begin{figure}[t]
    \centering
    \includegraphics[width=\columnwidth]{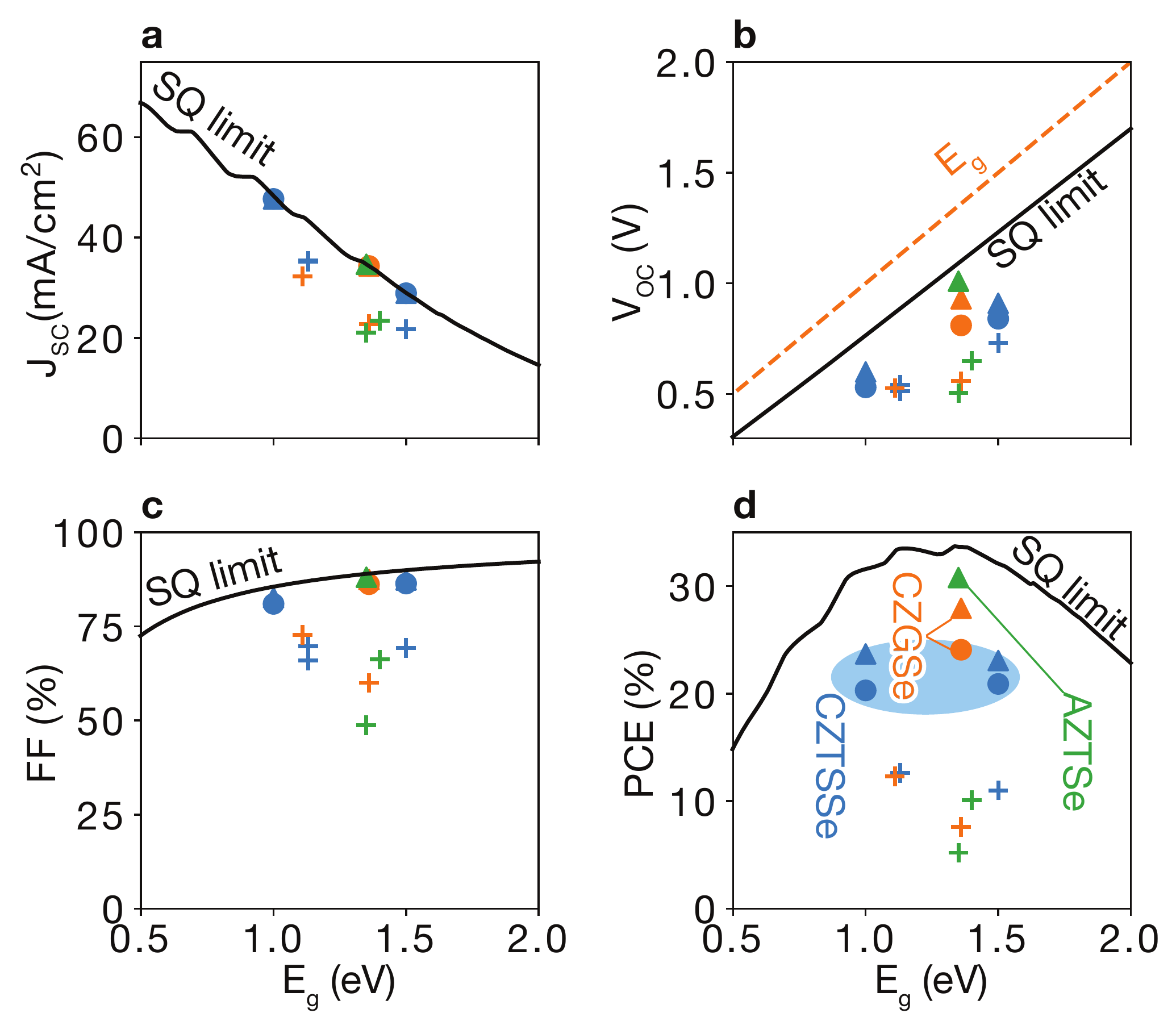}
    \caption{
    \textbf{Shockley-Queisser limit and trap-limited-conversion efficiency.} 
    \textbf{a}, Short-circuit current density $J_\text{SC}$, 
    \textbf{b}, open-circuit voltage $V_\text{OC}$,
    \textbf{c}, fill factor $FF$, and
    \textbf{d}, efficiency $\eta$.
    Filled symbols represent the trap-limited conversion (TLC), while a black line is the SQ limit.
    TLCs with doping (triangles) show better performances as compared to TLCs without doping (circles).
    Plus signs indicate experimental data for kesterite solar cells taken from
    Ref. \cite{Wang:2013gs,Kim:2016jr,Gershon:2016kt,Hadke:2018hx,Yan:2018dw,Choubrac:2018ex,Son:2019hm}
    }
    \label{fig:eff}
\end{figure}

\subsubsection*{Equilibrium defect concentrations}

\emph{Phase diagram:}
The growth environment of a crystal including elemental ratio, partial pressures, and temperature determines
the properties of the material including concentrations of the native defects.
In a theoretical framework, the growth conditions can be expressed using the thermodynamic chemical potential $\mu$ of each element.
We compare the energies of kesterites and their competing secondary phases,
showing a range of chemical potentials that favors the formation of kesterites,
using \texttt{CPLAP}.\cite{Buckeridge:2014ht}
We can avoid the formation of the secondary phases by a careful choice of synthesis conditions.
However even `pure' kesterites without secondary phases will contain native defects whose concentrations are controlled by this choice of chemical potentials. 

\emph{Formation energy of a defect:}
We calculated the formation energy $\Delta E_{f}(\ce{D}^q)$ of a defect $\ce{D}$ with the charge state $q$ as given by \cite{Freysoldt:2014ej}
\begin{equation}\label{eq:EForm}
\Delta E_{f}(\ce{D}^q) = E_\text{tot}(\ce{D}^q) - E_\mathrm{tot}(\mathrm{bulk}) - \sum_i N_i \mu_i + qE_F + E_\mathrm{corr},
\end{equation}
where $E_{tot}(\mathrm{bulk})$ and $E_{tot}(\ce{D}^q)$ are the total energies of a bulk supercell and a supercell containing the defect $\ce{D}^q$, respectively.
In the third term on the right-hand side, 
$N_i$ is the number of atoms $i$ added to the supercell, and 
$\mu_i$ is its chemical potential which is limited by the aforementioned phase diagram.
$E_{F}$ is the Fermi level, and $E_\mathrm{corr}$ is a correction term to 
account for the  spurious electrostatic interaction due to periodic boundary conditions.\cite{Freysoldt:2009ih,Kumagai:2014ih}

\emph{Self-consistent Fermi level:}
For a given synthesis condition (set of atomic chemical potentials), the formation energy is a function of the Fermi level as shown in Eq. \ref{eq:EForm}, 
while the Fermi level is determined by the concentrations of charged defects and carriers.
Thus we calculate the equilibrium concentrations of defects and carriers, and the Fermi level self-consistently
under the constraint of charge neutrality condition for overall system of defects and charge carriers using \code{SC-FERMI}\cite{Buckeridge:2019hd}. 

For a given Fermi level, the equilibrium concentration of a defect $N(\ce{D}^q)$ is given by
\begin{equation}\label{eq:concnt}
    N(\ce{D}^q) = N_\mathrm{site} g \mathrm{e}^{-\flatfrac{\Delta E_{f}(\ce{D}^q)}{k_B T}},
\end{equation}
where $N_\text{site}$ and $g$ are the number of available sites per unit volume and the degeneracy of the defect, respectively.
In the dilute limit, the competition between defects is negligible.
The partition function is approximated as 1 (i.e. the majority of lattice sites are regular).
Note that we use the internal energy of formation to calculate the defect density, neglecting the vibrational entropy change.
Thus the estimated defect densities are lower bounds.\cite{Walsh:2011ju}

The concentrations of holes $p_0$ and electrons $n_0$ are determined by the effective density of states of valence band $N_\mathrm{V}$ and conduction band $N_\mathrm{C}$:
\begin{equation}
\begin{split}
\label{eq:ss_carr_conct}
p_0 &= N_\mathrm{V} \mathrm{e}^{-\flatfrac{E_{F} - E_\mathrm{VBM}}{k_B T}},\\ 
n_0 &= N_\mathrm{C} \mathrm{e}^{-\flatfrac{E_\mathrm{CBM} - E_{F}}{k_B T}}. 
\end{split}
\end{equation}
Here, $E_\mathrm{VBM}$ and $E_\mathrm{CBM}$ are the reference energies of the valence band maximum and conduction band minimum, respectively.

The net charge of defects should be compensated by the net charge of electrons and holes:
\begin{equation}\label{eq:neutral}
    \sum_{i,j} q_{j} N({\ce{D}^{q_j}_i}) = p_0 - n_0.
\end{equation}
Thus, we iteratively update the Fermi level until the charge neutrality condition (Eq. \ref{eq:neutral}) is satisfied.
First, we determined the equilibrium concentration of defects at high temperature ($T_{an}=\SI{800}{\kelvin}$) and equilibrated their charge states at room temperature ($T_{op}=\SI{300}{\kelvin}$) with a fixed concentration of defects.

\subsubsection*{Defect levels}
A defect can change its charge state by capturing or emitting carriers.
The recombination process requires that defects are electrically active with more than one charge state.
The energy required to change the charge state of the defect level is often referred to as a \emph{thermal} activation energy or a charge-transition-level.
In modern defect theory, the defect level $\ce{D}$ is calculated as the position of Fermi level where the formation energies with two charge states of $q_1$ and $q_2$ are equal:
\begin{equation}\label{eq:E_T}
E_{T}(q_1/q_2; \ce{D}) = \frac{\Delta E_{f}(E_F = 0; \ce{D}^{q_1}) - \Delta E_{f}(E_F = 0; \ce{D}^{q_2})}{q_2 - q_1}.
\end{equation}

\begin{figure}[t]
    \centering
    \includegraphics[width=\columnwidth]{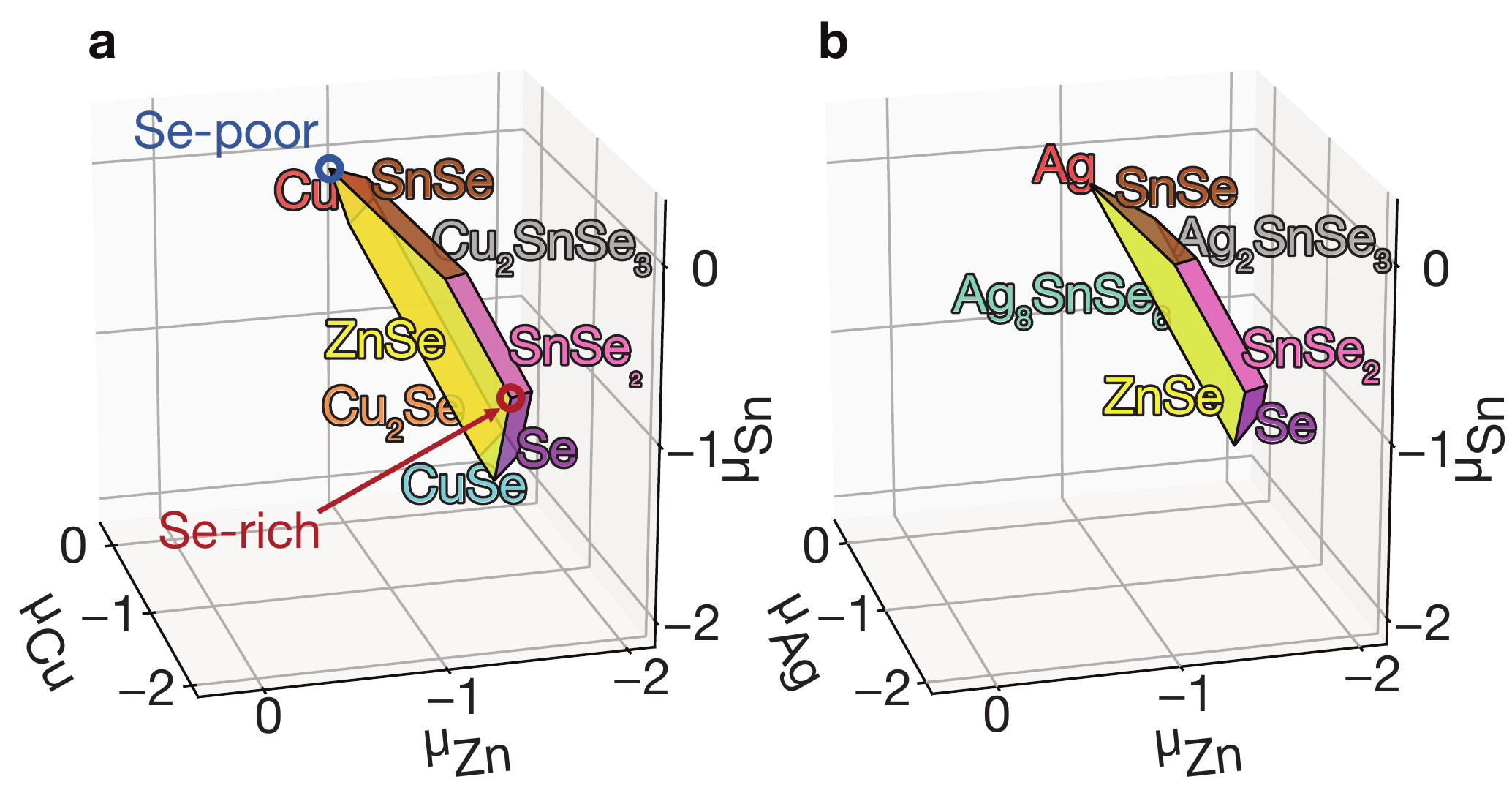}
    \caption{
    \textbf{Growth condition.}
    Calculated phase diagrams of \ce{Cu2ZnSnSe4}  (\textbf{a}) and  \ce{Ag2ZnSnSe4} (\textbf{b})
    where $\mu_{i}=0$ represents the chemical potential of element $i$ in its elemental state.
    Each plane represents a phase boundary with the secondary phase.
    Blue and red circles indicate Se-poor and Se-rich conditions, respectively.
    }
    \label{fig:pd}
\end{figure}

\subsubsection*{Carrier capture coefficient}

Nonradiative carrier capture via a defect is triggered by a  vibration and the associated electron-phonon coupling between the localised trap state and the delocalised free carriers.
The initial excited state, for example, a positively charged donor (\ce{D^+}) with an electron in the conduction band ($\mathrm{e}^-$), vibrates around the equilibrium geometry.
The deformation of the structure causes the electronic energy level of the trap state to oscillate.
As the energy level approaches the conduction band, the probability for the defect to capture the electron increases significantly.
When the electron is captured, the donor becomes neutral $\ce{D}^0$ and relaxes to a new equilibrium geometry by emitting multiple phonons.
To describe and predict such a process, 
quantitative accounts of the electronic and atomic structures, as well as vibrational properties of the defect are essential.

The carrier capture coefficient $C$ can be expressed 
using the electron-phonon coupling $W_{ct}$
and the overlap of phonon wave functions $\mel{\xi_{cm}}{\Delta Q}{\xi_{tn}}$,\cite{Alkauskas:2014kk,Kim:2019hp}
which is given by 
\begin{equation}\label{eq:CC}
\begin{split}
    C = & \Omega g \frac{2\pi}{\hbar} \abs{W_{ct}}^2 
          \sum_{m,n} w_{m} \abs{\mel{\xi_{tn}}{\Delta Q}{\xi_{cm}}}^2 \\
        &\times \delta(\Delta E + \epsilon_{cm} - \epsilon_{tn})
\end{split}
\end{equation}
where $\Omega$ and $g$ denote the volume of supercell and  the degeneracy of the defect, respectively.
$\psi$ and $\xi$ are electron and phonon wave functions, respectively, and 
the subscripts $c$ and $t$ specify the free carrier and trap states.
In this formalism, the temperature-dependence is determined by the thermal occupation number $w_{m}$ of the initial vibrational state.
In the following discussion, we calculate the capture coefficients at room temperature.
We employ an effective configuration coordinate $\Delta Q$ for the phonon wave functions
and adopt static coupling theory for $W_{ct}$.
The Coulomb attraction and repulsion between charged defects and carriers are accounted for by the Sommerfeld factor.\cite{Passler:1976gi,Landsberg:2009uO} 
See Supplementary information for details.

\subsubsection*{Steady-state illumination}

Under illumination or bias voltage, the steady-state electron and hole concentrations deviate from those determined by the equilibrium Fermi level.
The amount of applied voltage $V$ is the difference between the electron and hole \textit{quasi}-Fermi levels ($E_{F,n}$ for electron and $E_{F,p}$ for hole) which are functions of an additional carrier concentration $\Delta n$:
\begin{equation}
\label{eq:V_n}
\begin{split}
    \mathrm{q}V(\Delta n) = & E_{F,n} (\Delta n) - E_{F,h} (\Delta n) \\
                 = & E_\mathrm{CBM} + k_B T \ln(\frac{n_{0} + \Delta n}{N_{C}}) \\
                 - & E_\mathrm{VBM} + k_B T \ln(\frac{p_{0} + \Delta n}{N_{V}}) \\
                = & E_g + k_B T \ln( \frac{(n_{0} + \Delta n)(p_{0} + \Delta n)}{N_{C}N_{V}}),
\end{split}
\end{equation}
where we ignore the voltage drop due to a series resistance and a shunt across the device.
One can rewrite Eq. \ref{eq:V_n} for $\Delta n$ as a function of $V$:
\begin{equation}
\begin{split}
    \Delta n (V) 
    = &  \frac{1}{2}\left[ 
    -n_0 -p_0 + 
    \sqrt{(n_0 + p_0)^2     - 4 n_i^2 \left( 1- \mathrm{e}^\frac{\mathrm{q}V}{k_B T}\right)  } \right],
\end{split}
\end{equation}
where $n_i^2 = n_0 p_0 = N_C N_V \mathrm{e}^\frac{-E_{g}}{k_B T}$.
Accordingly, the steady-state concentrations of electron $n$ and hole $p$ under applied voltage $V$ are given by
\begin{equation}
\begin{split}\label{eq:npV}
    n(V) &= n_0 + \Delta n(V),\\
    p(V) &= p_0 + \Delta n(V).
\end{split}
\end{equation}

\subsection{Trap limited conversion efficiency}
By taking into account the carrier annihilation due to both radiative recombination (Eq. \ref{eq:SQJV}) and nonradiative recombination (Eq. \ref{eq:SRH}),
the trap-limited current density $J$ under a bias voltage $V$ is given by
\begin{equation}\label{eq:jv}
\begin{split}
    J(V; W) =& J_\mathrm{SC}(W) + J_\mathrm{0}^\mathrm{rad}(W)(1 - \mathrm{e}^{\frac{\mathrm{q}V}{k_B T}})\\
            &- q R_\mathrm{SRH}(V) W.
\end{split}
\end{equation}
The voltage-dependent nonradiative recombination rate $R_\text{SRH}$ is obtained by combining Eq. \ref{eq:SRH}, \ref{eq:concnt}, \ref{eq:E_T}, \ref{eq:CC}, and \ref{eq:npV}.
Finally, we evaluate the photovoltaic maximum efficiency:
\begin{equation}
    \eta = \text{max}_V \left(\frac{JV}
    {q \int_{0}^{\infty} E \Phi_\text{sun}\left( E \right)  \dd{E} } \right).
\end{equation}



\begin{figure*}[]
    \centering
    \includegraphics[width=2\columnwidth]{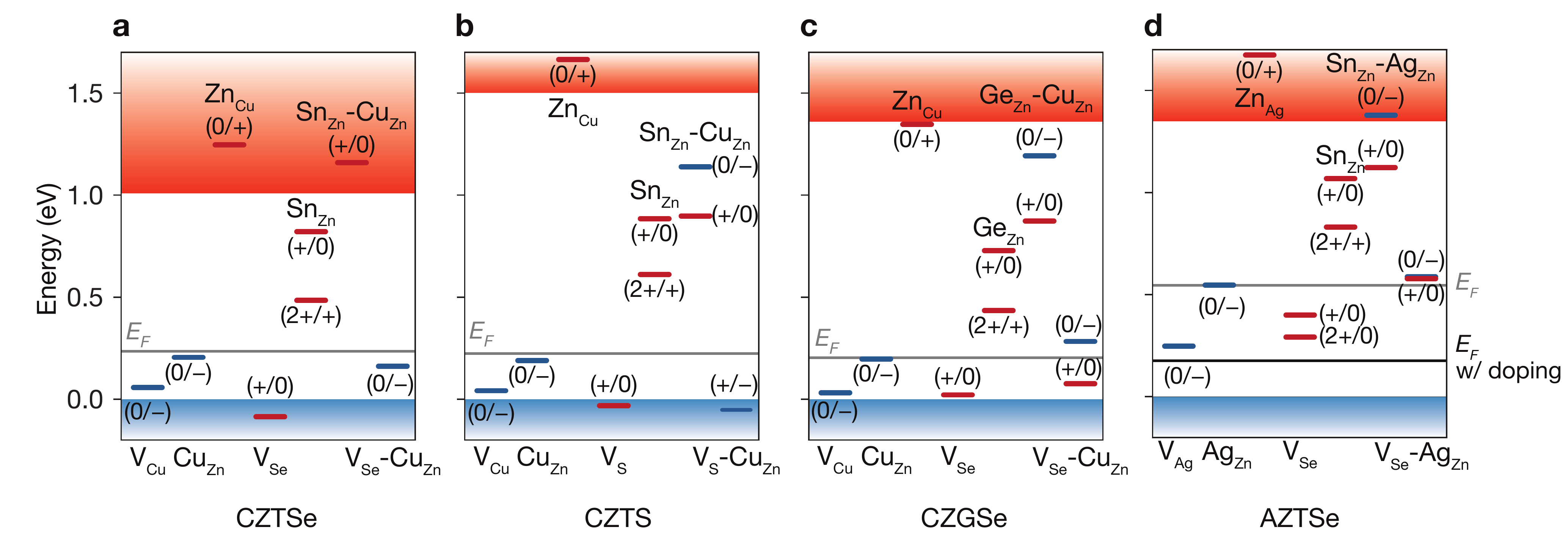}
    \caption{
    \textbf{Defect levels of native defects.}
    Donor (red) and acceptor (blue) levels of native point defects of \ce{Cu2ZnSnSe4} (\textbf{a}), \ce{Cu2ZnSnS4} (\textbf{b}), \ce{Cu2ZnGeSe4} (\textbf{c}), and \ce{Ag2ZnSnSe4} (\textbf{d}).
    Blue and red bands represents valence and conduction bands, respectively.
    Fermi levels are shown in gray lines.
    The black line in \textbf{d} represents the Fermi level of \ce{Ag2ZnSnSe4} with a doping density of \SI{e20}{cm^{-3}}.
    }
    \label{fig:E_level}
\end{figure*}

\section{Results}
We apply our scheme to kesterite solar cells (\ce{Cu2ZnSnSe4}, \ce{Cu2ZnSnS4}, \ce{Cu2ZnGeSe4}, and \ce{Ag2ZnSnSe4}),
with details presented in the Methods section and Supplementary Table 1.

\subsection{\ce{Cu2ZnSnSe4} and \ce{Cu2ZnSnS4}}
\emph{Shockley-Queisser limit:}
In the SQ limit under 1-sun (AM1.5g) illumination, 
the maximum efficiency of CZTSe with a band gap of 1 eV is 31.6\% (see Fig. \ref{fig:eff}) with a \VOC~ of 0.77 V.
Next, we calculate the nonradiative recombination rate due to native defects.

\emph{Growth conditions:}
Single-phase CZTSe is formed when the chemical potential of the elements are in the phase field of CZTSe as shown in Fig. \ref{fig:pd}a.
The phase diagram of CZTSe has a small volume with a narrow window of available chemical potentials,
which the stability of \ce{ZnSe} is largely responsible for.
At high Zn-ratio, Zn atoms tend to form \ce{ZnSe} rather than to incorporate at their lattice sites in CZTSe.
Later, we will show that this poor incorporation of Zn results in high concentrations of antisite defects:
\ce{Cu_{Zn}} and \ce{Sn_{Zn}}, which are responsible to the \textit{p}-type Fermi level and the low carrier lifetime, respectively.

\emph{Defect levels:}
Point defects introducing defect levels close to the band edge are categorized as shallow and generate free carriers.\cite{Park:2018et}
On the other hand, deep defects are often responsible for carrier trapping and nonradiative recombination, limiting the efficiency of solar cells.\cite{Park:2018et}

The band structure of \ce{CZTSe} is composed of antibonding Sn 5\textit{s}-Se 4\textit{p}$^{*}$ state at the lower conduction band and antibonding Cu 3\textit{d}-Se 4\textit{p}$^{*}$ state at the upper valence band.
According to models for defect tolerance,\cite{Zakutayev:2014dxa, Walsh:2017ko}
the Cu dangling bond would produce a shallow level, 
while a deep level can be introduced by the Sn dangling bond.
Moreover, the cation antisites, especially \ce{(Cu{,}Zn)_{Sn}} and \ce{Sn_{(Cu{,}Zn)}} are expected to be deep due to the large difference in the site electrostatic (Madelung) potentials.\cite{Kim:2018jd}

\begin{figure}[b]
    \centering
    \includegraphics[width=\columnwidth]{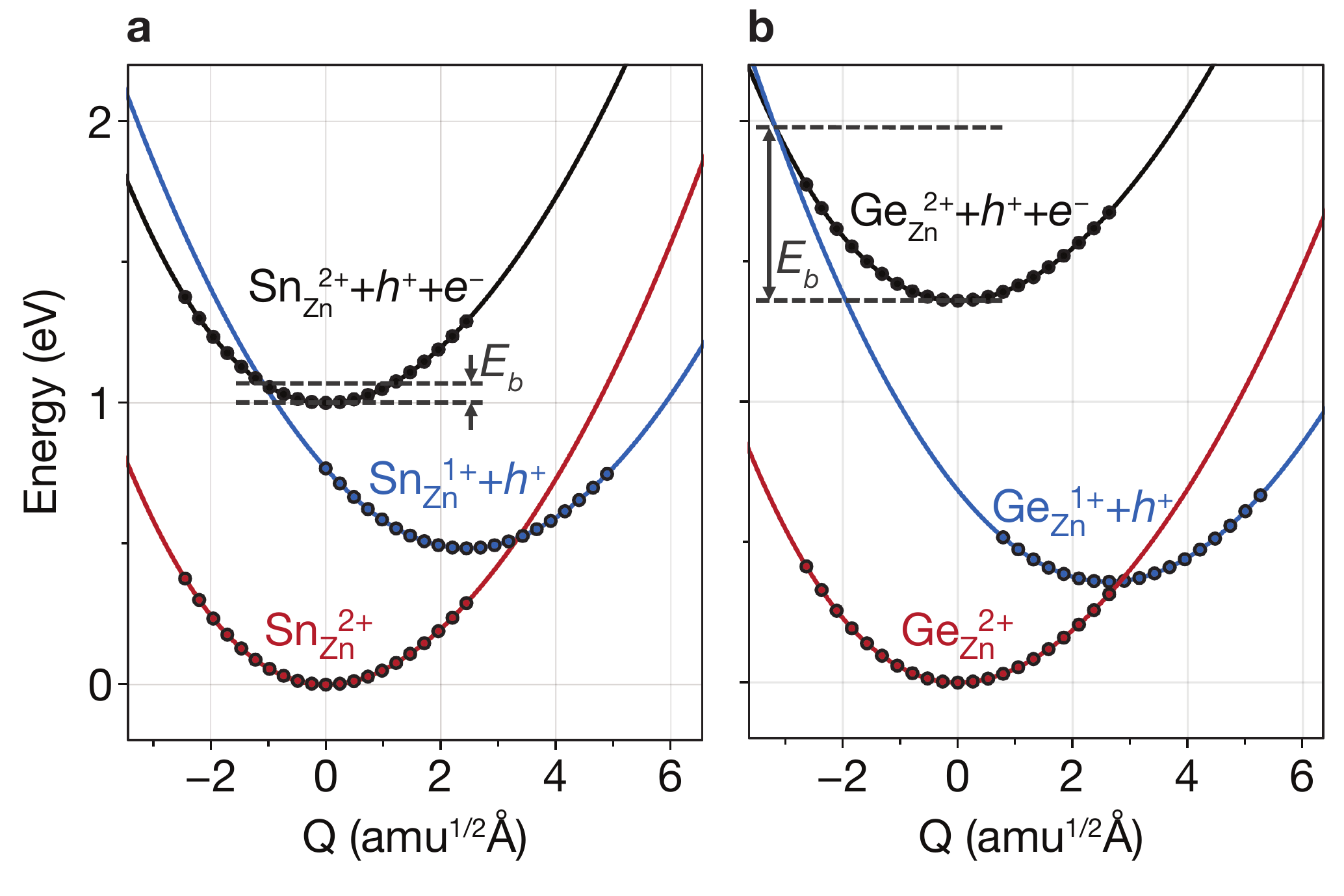}
    \caption{
    \textbf{Configuration coordinate diagram for carrier capture.} 
    Potential energy surfaces for the vibrations of \ce{Sn_{Zn}} (2+/+) in \ce{Cu2ZnSnSe4} (\textbf{a}) 
    and \ce{Ge_{Zn}} (2+/+) in \ce{Cu2ZnGeSe4} (\textbf{b}).
    The solid circle represents the relative formation energy calculated using DFT, 
    and the line is a spline fit.
    $E_{b}$ represents the electron-capture barrier.
    }
    \label{fig:E_CC}
\end{figure}

Admittance spectroscopy (AS) measurements identified several shallow  acceptors in \ce{Cu2ZnSn(S{,}Se)4}, CZTSSe, CZTSe and CZTS at an energy range between 0.05-0.17 eV   \cite{Gunawan:2012kx,Khadka:2015jx,Levcenko:2016hw,Koeper:2017fu,Yang:2018hq,Yang:2019do}, which were attributed to \ce{V_{Cu}} and \ce{Cu_{Zn}}.
They also found a deep level close to the midgap ($E_T = ~$\SI{0.5}{eV}).
A series of deep-level transient spectroscopy (DLTS) experiments also revealed the presence of the shallow levels as well as a broad spectrum of deep levels around the mid gap. \cite{Li:2013bxa,Das:2014kyb,Kheraj:2016hw}
Transient photocapacitance (TPC) spectra showed sub-band-gap absorption \emph{via} deep defects near \SI{0.8}{eV} with broad bandwidth.\cite{Miller:2012cs,Islam:2015bh}
Theoretical calculations\cite{Han:2013bg,Kim:2018jd,Kim:2019iv,Li:2019bx} revealed the atomic origins of shallow defects:
acceptors \ce{V_{Cu}} and \ce{Cu_{Zn}} and a donor \ce{Zn_{Cu}}.
Several atomic models for the deep defects have been proposed such as 
\ce{(Cu_3)_{Sn}}, 
\ce{Sn_{Zn}},
\ce{V_{S}}, \ce{V_{S}}-\ce{Cu_{Zn}}, and \ce{Sn_{Zn}}-\ce{Cu_{Zn}}.\cite{Han:2013bg,Kim:2018jd,Kim:2019iv,Li:2019bx}

First, we find shallow acceptors (\ce{V_{Cu}} and \ce{Cu_{Zn}}) and a shallow donor (\ce{Zn_{Cu}}) (see Fig. \ref{fig:E_level}a and Supplementary Table 2).
Due to the similar ionic radii of \ce{Cu} and \ce{Zn}, 
the energy cost for the formation of \ce{Cu_{Zn}} and \ce{Zn_{Cu}} is very low.
The very low formation energy of \ce{Cu_{Zn}} for every set of chemical potentials is largely responsible for the \textit{p}-type Fermi level around 0.2 eV.
We find that the decrease in oxidation state of Sn found in \ce{V_{Se}}, \ce{Sn_{Zn}} and \ce{V_{Se}}-\ce{Cu_{Zn}} produces deep levels, 
similar to those found in CZTS.\cite{Han:2013bg, Kim:2018jd, Kim:2019iv, Li:2019bx}
The deep donor \ce{Sn_{Zn}} becomes shallow when it combines with \ce{Cu_{Zn}} because of the Coulomb attraction between the ionized donor and acceptor.\cite{Kim:2019iv}

\emph{Capture coefficients:}
As Cu-based kesterites are intrinsic \textit{p}-type semiconductors, 
the carrier lifetime is determined by the electron-capture processes via deep defects.
We calculate electron-capture coefficients of the selected deep defects:
\ce{V_{Se}}-\ce{Cu_{Zn}} and \ce{Sn_{Zn}},
satisfying the criterion
$E_\textrm{CBM} - E_T >  E_\textrm{VBM} - E_{F} + 0.1$\si{eV} so that $n_t \ll p$ at $T= $\SI{300}{\K}, 
and $N_T>$\SI{e14}{cm^{-3}}.

Due to the Sn reduction associated with these defects, they exhibit not only a deep level, but also a large structural relaxation that leads to large electron-capture coefficients.\cite{Kim:2018jd, Kim:2019iv}
Fig. \ref{fig:E_CC}a shows the configuration coordinate for \ce{Sn_{Zn}}(2+/1+),
illustrating that the carrier-capture barrier is small due to the large lattice relaxation,
the horizontal shift of the potential energy surface of \ce{Sn^{1+}_{Zn}} with respect to that of \ce{Sn^{2+}_{Zn}}.
Thus, we find that \ce{Sn_{Zn}}(2+/1+) has a large electron-capture coefficient of $\SI{9E-7}{cm^3s^{-1}}$ 
(corresponding to the capture cross section of $\SI{9.29E-14}{cm^3s^{-1}}$), which classify them as \emph{killer} centers.\cite{Stoneham:gb}
Note that the minority-carrier capture coefficient of these \emph{native} defects in CZTSe are of a similar order of magnitude of 
the most detrimental \emph{extrinsic} impurities in Si solar cells.\cite{Macdonald:2004eh,Peaker:2012hd}
We also find a large electron-capture coefficient of \ce{V_{Se}}-\ce{Cu_{Zn}},
which is listed in Supplementary Table 2. 

\emph{Equilibrium concentration:}
The concentration of native point defects can be tuned through the chemical environment.
However, we find that it is difficult to reduce the concentration of the \emph{killer} centers in CZTSe.
For example, to reduce the concentration of \ce{Sn_{Zn}},
we need: i) to increase Zn incorporation, ii) to decrease Sn incorporation, or iii) to decrease hole concentration.
These are difficult to achieve due to the narrow thermal equilibrium phase diagram.
First, the high-Zn incorporation is difficult to achieve because of the aforementioned high stability of \ce{ZnSe}.
On the other hand, the incorporation can be tuned to decrease the concentration of \ce{Sn_{Zn}}.
The low Sn incorporation, together with the low Zn incorporation, will, however, result in the formation of the highly conductive secondary phases of \ce{CuSe} and \ce{Cu2Se} (see Fig. \ref{fig:pd}a), which can electrically short the device.\cite{Dimitrievska:2016cj}
Thus, the low Sn incorporation should actually be avoided. 
We also find the hole concentrations are high under all conditions
due to the high concentrations of \ce{Cu_{Zn}}, which is also the consequence of the poor Zn incorporation.
Therefore, it is difficult to decrease the concentrations of \ce{Sn_{Zn}} in thermal equilibrium.

\begin{figure}[]
    \centering
    \includegraphics[width=\columnwidth]{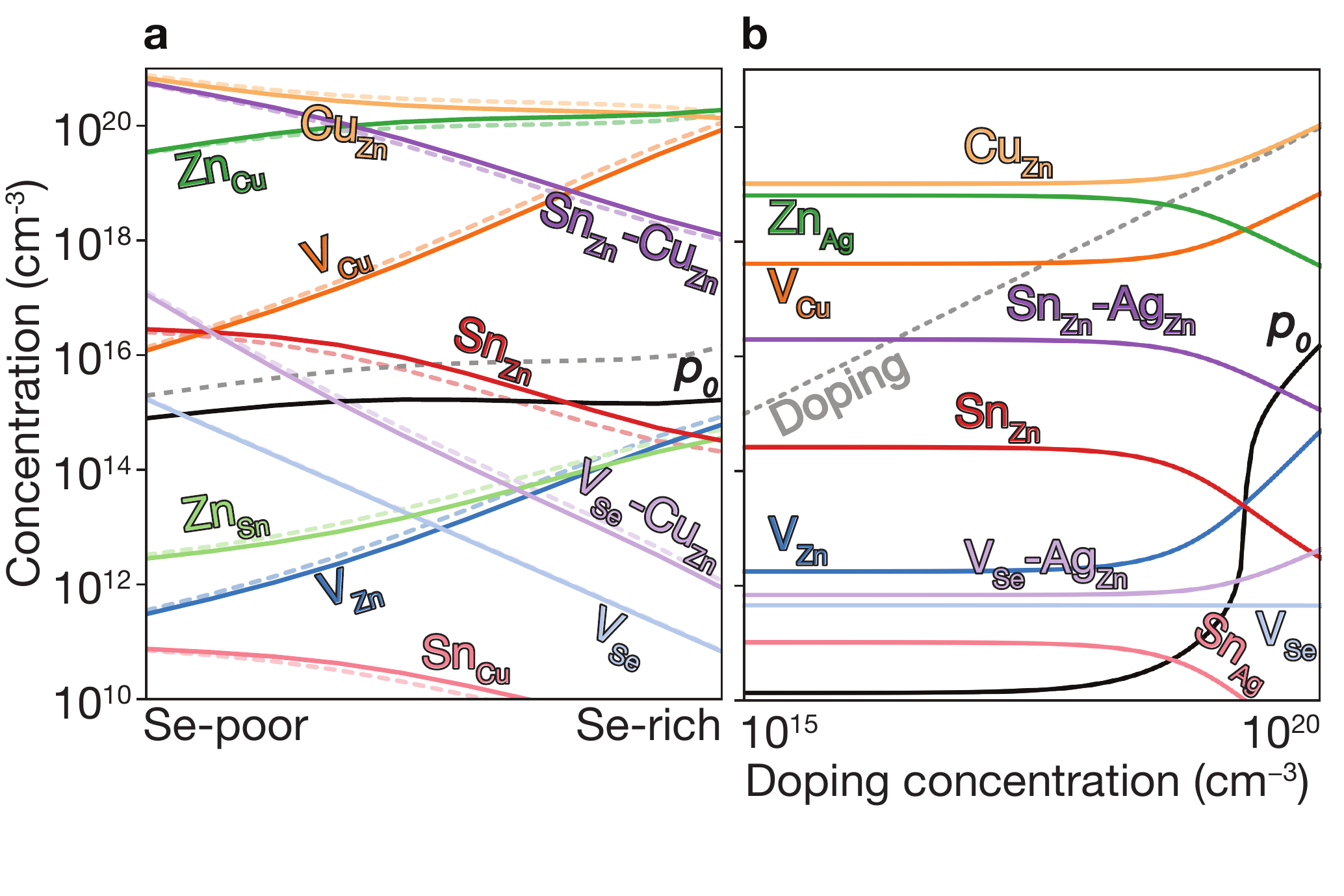}
    \caption{
    \textbf{Concentrations of native defects.}
    \textbf{a}, The concentrations of native defects in CZTSe.
    The dashed lines represent the concentration with the doping during the growth (see text for details).
    \textbf{b}, The concentrations of native defects in AZTSe with the doping during the growth.
    The dashed diagonal line represents the doping concentration.
    }
    \label{fig:concnt}
\end{figure}

Fig. \ref{fig:concnt}a shows the equilibrium concentrations of native defects under Se-poor and Se-rich conditions (see Fig. \ref{fig:pd}a).
Under Se-poor conditions, we find high concentration of \ce{V_{Se}}-\ce{Cu_{Zn}}, which is an efficient recombination center.
While their concentrations can be significantly decreased through Se incorporation,
the concentration of \ce{Sn_{Zn}} can not be decreased below \SI{e14}{\per\cubic\centi\metre},
which limits the maximum performance of CZTSe solar cells.

Finally, we stress that the capture cross section and defect concentrations of the dominant recombination center in CZTSe (\ce{Sn_{Zn}}) are in good agreement with experiments.\cite{Levcenko:2016hw,Hages:2017co}
Our previous admittance spectroscopy \cite{Levcenko:2016hw} revealed a deep defect level located at \SI{0.5}{eV}.
Based on the thermal emission prefactors of up to \SI{5e12}{\cm\per\s} at room temperature,
we estimate the capture cross section as \SI{1e-13}{\cm^2} which agrees well with our calculation of \SI{9e-14}{\cm^2} (see Supplementary Table 2).
We also find the longest minority-carrier lifetime achievable is less than \SI{5.5}{\nano\s} in CZTSe 
which closely agrees with the previous assessment of the \emph{real} minority-carrier lifetime of below \SI{1}{\nano\s} based on time-resolved photoluminescence.\cite{Hages:2017co,li2019relating}

\begin{figure}[b]
    \centering
    \includegraphics[width=\columnwidth]{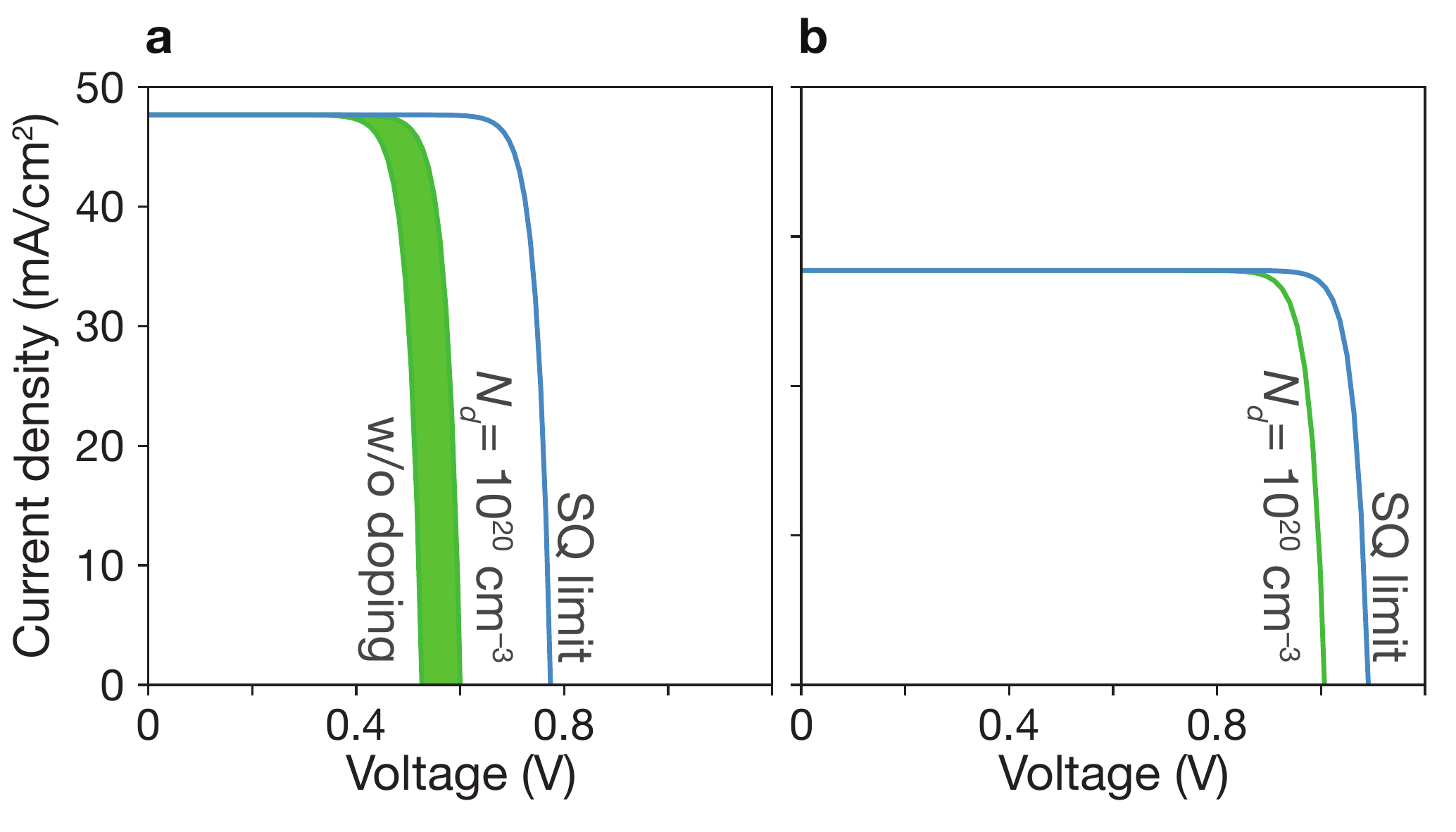}
    \caption{
    \textbf{Current-voltage simulation.}
    $J$-$V$ curves for CZTSe (\textbf{a}) and AZTSe (\textbf{b}) solar cells
    based on the properties of the bulk absorber materials and not including inferfacial processes.
    Green lines represent the TLCs with various doping concentrations up to \SI{e20}{\cm^{-3}}.
    The SQ limit is shown in the blue curve. 
    }\label{fig:sim}
\end{figure}

\emph{Trap limited conversion efficiency:}
We calculate the current-voltage characteristic (Eq. \ref{eq:jv}) of a CZTSe solar cell containing the equilibrium concentrations of native point defects under the Se-rich condition (See Fig. \ref{fig:sim}a).
We used the a film thickness of $\SI{2}{\micro\meter}$.
The overall power-conversion efficiency is 20.3\%, which is below two thirds of the SQ limit of 31.6\% (see Fig. \ref{fig:eff} and Table \ref{tab:eff}).

\emph{Sulfide kesterite:}
\ce{Cu2ZnSnS4} (CZTS) also suffers from nonradiative recombination 
due to the redox activity of Sn and the narrow phase space limited by the high stability of \ce{ZnS}.
Similar to \ce{Sn_{Zn}} in CZTSe, we find the large structural relaxation for \ce{Sn_{Zn}} that causes fast carrier capture.
Moreover, although the defect complex \ce{Sn_{Zn}}-\ce{Cu_{Zn}} is a shallow donor in CZTSe,
in CZTS having the larger band gap of \SI{1.5}{eV}, \ce{Sn_{Zn}}-\ce{Cu_{Zn}} produces the deep donor level at $E_{T} = 0.90$ eV 
as shown in Fig. \ref{fig:E_level}a and b.
Thus, the recombination pathways in CZTS are not only through the isolated \ce{Sn_{Zn}} but also the \ce{Sn_{Zn}} bound to the acceptor \ce{Cu_{Zn}}, which agrees well with a previous theoretical study \cite{Li:2019bx}.
We find that the similar behavior for \ce{Ge_{Zn}} in \ce{Cu2ZnGeSe4} which will be discussed in detail in the following subsection.
We calculate a nonradiative \VOC~ loss of 0.39 V, corresponding to an achievable \VOC~ of 0.84 V and 
a maximum TLC of 20.9\% for CZTS, which is similar to that of CZTSe.

\subsection{\ce{Cu2ZnGeSe4}}
As the redox activity of Sn is one culprit that reduces the voltage and efficiency of CZTSe and CZTS devices,
we can suppress the nonradiative recombination 
by substituting Sn with other cations such as Si with a more stable $4+$ oxidation state. 
However, the SQ limit of \ce{Cu2ZnSiSe4} is below 16\% because of its large band gap of 2.33 eV.\cite{Shu:2013ed}
On the other hand, \ce{Cu2ZnGeSe4} (CZGSe) has an optimal band gap of 1.36 eV with an SQ limit of 33.6\%.
However, we find that the similar redox activity of Ge in CZGSe causes significant nonradiative recombination and limits the \VOC.

Ge also exhibits an inert-pair effect with large ionisation energy for the 4\textit{s} orbital.  
Thus, Ge-related defects (\ce{Ge_{Zn}}, \ce{Ge_{Zn}}-\ce{Cu_{Zn}}, \ce{V_{Se}} and \ce{V_{Se}}-\ce{Cu_{Zn}}) introduce deep donor levels in the band gap. 
\ce{Ge_{Zn}} exhibits the similar potential energy surfaces to those of \ce{Sn_{Zn}} in CZTSe (Fig. \ref{fig:E_CC}b).
However, \ce{Ge_{Zn}} has a deeper donor level than that of \ce{Sn_{Zn}} due to the larger band gap of CZGSe (see supplementary Table 1).
As shown in Fig. \ref{fig:E_CC},
because the electron-capture processes due to \ce{Sn_{Zn}} and \ce{Ge_{Zn}} are in the so-called ``Marcus inverted region",\cite{Marcus:1985ev}
the deeper donor level of \ce{Ge_{Zn}} results in a higher energy barrier for electron-capture (0.62 eV).
We find a several orders of magnitude smaller electron-capture coefficient for \ce{Ge_{Zn}} (2+/1+) as compared to that of \ce{Sn_{Zn}} (2+/1+), 
implying that the recombination due to the isolated \ce{Ge_{Zn}} is unlikely to happen (see Supplementary Table 2).

However, the nonradiative recombination rate in CZGSe is still high due to defect complexation.
The abundant acceptor \ce{Cu_{Zn}} tends to form a defect complex with donors such as \ce{Ge_{Zn}}.
The Coulomb attraction between the ionized donor and acceptor further promote the formation of the complex.
Moreover, the donor-acceptor complex makes the defect level shallower ($E_{T}=\SI{0.87}{eV}$).\cite{Kim:2019iv}  
We find that the electron-capture barrier is \SI{71}{meV} for \ce{Ge_{Zn}}-\ce{Cu_{Zn}} (1+/0), which is the dominant recombination pathway in CZGSe. 
Although, we considered only the \ce{Ge_{Zn}} and \ce{Cu_{Zn}} pair bound at the closest site, 
in reality, there are a variety of complexes with a wide range of distances between \ce{Sn_{Zn}} and \ce{Cu_{Zn}}.
Such a spectrum of complexes are partially responsible for the broad defect levels in kesterites
measured in photocapacitance spectroscopies.\cite{Miller:2012cs,Islam:2015bh}

By taking into account the formation of defect complexes, 
we find significant nonradiative loss in CZGSe.
The maximum efficiency is predicted to be $21.9$\% with large non-radiative open-circuit voltage loss of 0.29 V (see Fig. \ref{fig:eff} and Table \ref{tab:eff}).

\subsection{Hydrogen and alkali-metal doping, and \ce{Ag2ZnSnSe4}}

\begin{figure}[t]
    \centering
    \includegraphics[width=\columnwidth]{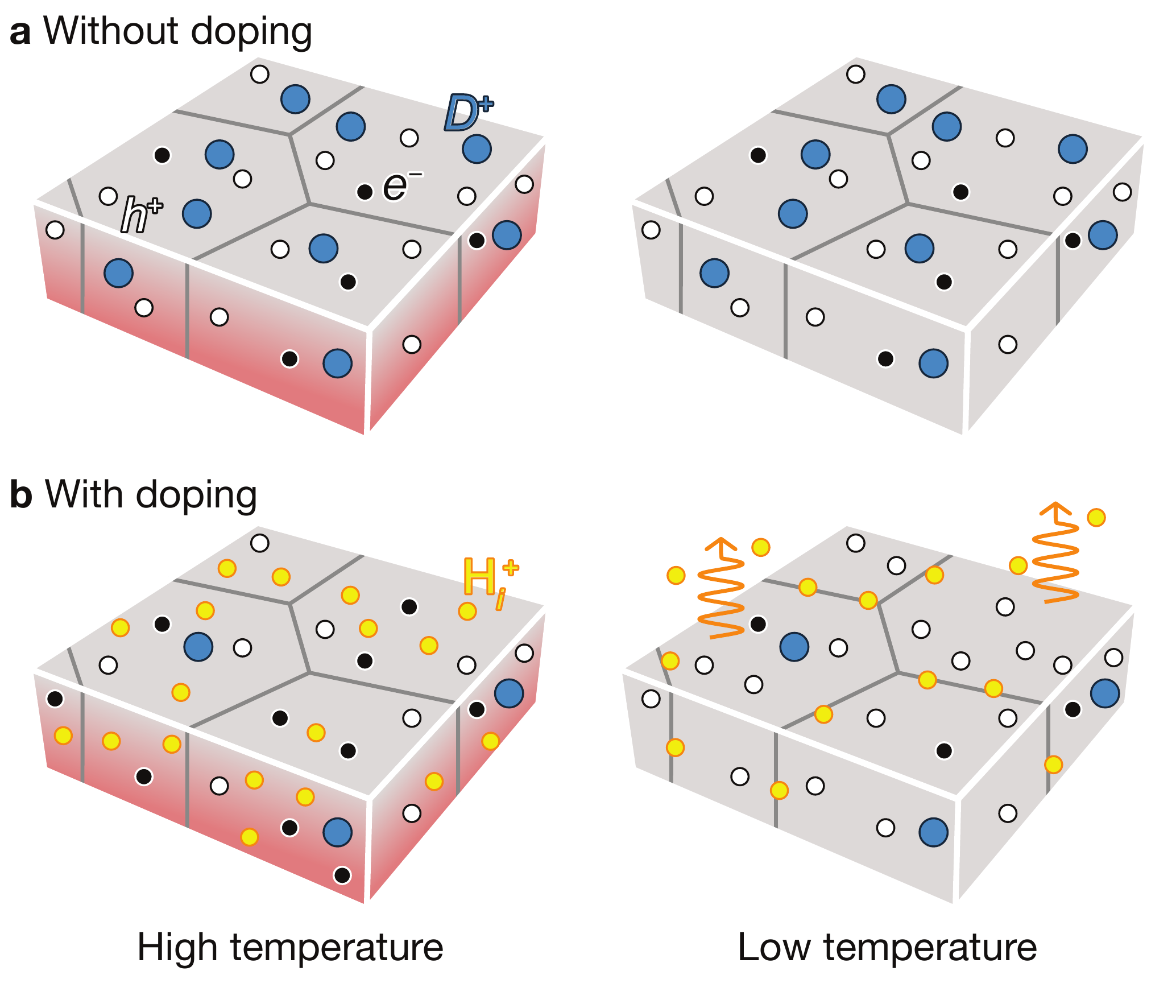}
    \caption{
    \textbf{The effect of hydrogen/alkali-metal doping on kesterites.} 
    Schematics for the formation of defects without doping (\textbf{a}) and with doping (\textbf{b}).
    During thermal annealing, the native defects are formed at high temperature (left panel),
    whose populations remain the same when the sample is cooled down to low temperature (right panel).
    A high concentration of hole (white circle) promote the formation of donors (blue circle).
    Dopants are marked as yellow circles.
    For the clarity, the acceptors are not drawn.
    }
    \label{fig:doping}
\end{figure}

As an additional lever to tune the defect profiles, 
we consider extrinsic doping.
The formation energy, and hence concentration, of a defect
depends on the chemical potential of an electron (Fermi level).
In CZTSe, CZTS, and CZGSe, the intrinsic Fermi levels are pinned $\sim$0.2 eV above the valence band (Fig. \ref{fig:E_level}a, b, and c),
promoting the formation of deep donors.
As illustrated in Fig. \ref{fig:doping}a,
such high concentrations of donors arise at high (growth) temperature and remain after cooling
because they are mostly immobile vacancies and antisites.
While \textit{n}-type doping can increase the Fermi level,
this type of doping will not increase the \VOC~ (efficiency) for a material with limited minority carrier lifetime,
because \textit{n}-type doping will decrease the \textit{p}-type conductivity.

Instead, we predict that hydrogen and alkali-metal doping is helpful to increase the efficiency.
At high temperature during the thin-film growth or thermal annealing,
the incorporation of the hydrogen or alkali metals at the interstitial sites will increase the Fermi level as they act as donors in \textit{p}-type semiconductors.\cite{Varley:2018dma}
The high Fermi level decreases the hole concentration and the formation of donor type defects as well (see Fig. \ref{fig:doping}c).
Since hydrogen and alkali-metals are mobile,
they tend to diffuse easily and segregate to the grain boundary or outgas,
when the thin-film cools down to the room temperature (see Fig. \ref{fig:doping}d).
The final thin-film will exhibit an increased hole concentration and longer carrier lifetime,
consistent with the experiments.\cite{Haass:2017da}
This is indeed the mechanism behind the success of hydrogen-codoping in nitride semiconductors.\cite{Nakamura:2015uu, Fioretti:2017bk}

We calculate the concentrations of defects in CZTSe with a \textit{n}-type doping concentration of $\SI{e20}{\per\cubic\centi\meter}$ at $T = T_{an}$.
Once the dopants are removed, 
the hole concentration increases by an order of magnitude at $T = T_{op}$, and the concentration of \ce{Sn_{Zn}} is significantly lowered (see Fig. \ref{fig:concnt}a).
Thus, the maximum efficiency increases up to $23.7$ \% (Fig. \ref{fig:eff} and Fig. \ref{fig:sim}a).
This requires a high level of doping to gain a noticeable improvement
due to the high concentration of native donors and acceptors, and the self-compensation mechanism via them.
Alkali-metal elements may be less effective dopants
due to their low solubility.\cite{Haass:2017da}
On the other hand, the previous calculations \cite{Varley:2018dma} have shown that the formation energies of \ce{H_i} in kesterites are low at \textit{p}-type Fermi-level,
suggesting high solubility of H in kesterites.
We also noted that
Son \textit{et al.} formed a S-Se grading in the current champion device \cite{Son:2019hm}
 using \ce{H2S} gas,
which may introduce the H-doping unintentionally and be responsible for the high efficiency.

The low formation energies and the high concentrations of \ce{Cu_{Zn}} and \ce{Zn_{Cu}} originate from the similar ionic radii of \ce{Cu^{1+}} and \ce{Zn^{2+}}.
We may decrease their concentrations by exploiting Ag substituting Cu or Cd substituting Zn.\cite{Gautam:2018jj}
\ce{Ag} substitution for \ce{Cu} gives \ce{Ag2ZnSnSe4} (AZTSe), which also has a narrow phase diagram as shown in Fig. \ref{fig:pd}b.
However, we find several orders of magnitude lower concentrations of the dominant acceptor and donor, \ce{Ag_{Zn}} and \ce{Zn_{Ag}} (see Fig. \ref{fig:concnt}b).
AZTSe is an intrinsic semiconductor under Se-rich conditions,
while \textit{n}-type Fermi level was found under Se-poor conditions.

For a set of atomic chemical potentials determined under Se-rich conditions, the calculated self-consistent Fermi-level is \SI{0.55}{eV} above the valence band.
Due to the low hole concentration in AZTSe, Eq. \ref{eq:SRHp} is not valid, and the hole-capture process becomes the bottleneck in the recombination process
owing to the high hole-capture barrier of \SI{0.20}{eV} as compared to the electron-capture barrier of \SI{0.11}{eV}.
However, due to the high Fermi level in AZTSe or even n-type conductivity, Ag-based solar cells based on the commonly used thin-film  architecture for Cu-based kesterites (Mo/kesterite/CdS/ZnO/ITO), have been found to exhibit limited device performance.\cite{Gershon:2016kt,Gershon:2017eh,Jiang:2019bj}  
Notwithstanding these practical challenges, we predict that Ag-based kesterites should show much lower non-radiative recombination and thus possess a significantly larger efficiency potential than the previously discussed Cu- or Ge-based kesterites.
Indeed, increased photoluminescence quantum yields (PLQY) have been recently observed for Ag-substituted kesterites.\cite{plAgKest}

An extrinsic \textit{n}-type doping level of $\SI{e20}{\per\cubic\centi\meter}$ during growth can lower the room temperature Fermi-level to \SI{0.18}{eV}.
As shown in Fig. \ref{fig:concnt}b, this causes the concentration of \ce{Sn_{Zn}} to decrease below $\SI{e14}{\per\cubic\centi\meter}$,
enhancing the maximum efficiency up to $30.8$\% (see Fig. \ref{fig:eff} and Fig. \ref{fig:sim}),
implying that co-doped AZTSe is a promising material as a \textit{p}-type absorber if the synthesis and processing be appropriately controlled.

\subsection{Calculation of optoelectronic parameters} 

The achievable solar cell parameters estimated for four types of kesterite materials using our first-principles approach are summarized in Table I, and compared with the (defect-free) Shockley-Queisser limit, as well as current champion devices. 

The Ge- and Ag-based materials so far significantly underperform, and that big leaps in efficiency appear possible by the proposed co-doping strategy. 
Device performance can be limited by a number of non-idealities such as non-optimised functional layers, 
wrong band line-ups, as well as interface recombination.
It is therefore helpful to consider the main (absorber layer) optoelectronic parameters that are experimentally accessible even without building devices. 
Among the most relevant to judge potential device performance are carrier lifetime, net doping density, and external PLQY, which indicates the ratio of radiative recombination over the total recombination, typically dominated by non-radiative processes. 
The PLQY can be estimated from non-radiative voltage loss using $\Delta V_\mathrm{OC}^\mathrm{nonrad} = k_B T \ln(\mathrm{PLQY})$.\cite{ross1967some} 

A summary of these parameters, calculated from first-principles, are listed in Table \ref{tab:par}, indicating small PLQY and lifetimes for CZTS and large PLQY and long lifetimes for co-doped AZTSe. 
The small predicted PLQY for CZTS is in agreement with observations that the luminescence yield of this material is consistently below the detection limit (ca. $\SI{1e-4}\%$). 
Also, the PLQY value of $\SI{1e-2}\%$ is consistent with recent reports of $\SI{1.5e-3}\%$ measured on a CZTSe single crystal\cite{li2019relating} and of $\SI{3e-3}\%$ on 11.6\% efficient Li-doped CZTSSe solar cells.\cite{cabas2018high} 

In these solar cells the lifetime did not change significantly with Li-doping, while the PLQY and net doping density increased, again inline with our predictions. 
With regards to the calculated minority carrier lifetimes, we point out that the small estimated lifetimes for CZTS and CZTSe are in good agreement with recent findings indicating that reported carrier lifetimes for kesterites are often overestimated and that (typical) real lifetimes are in fact below \SI{1}{ns}.\cite{Hages:2017co}

\begin{table}[ht]
    \centering
    \begin{tabular*}{\columnwidth}{lllllll}
           & $E_\mathrm{gap}$  & $\eta$  & \JSC  & \VOC       &  FF & Reference \\
           & (eV) & (\%) & ($\si{mA\per cm^{2}}$) & ($\si{V}$) & (\%) & \\
\hline
CZTS       & 1.50 & 32.1 & 28.9 & 1.23 & 90.0 & SQ limit \\ 
CZTSe      & 1.00 & 31.6 & 47.7 & 0.77 & 85.7 & SQ limit \\
CZGSe      & 1.36 & 33.3 & 34.3 & 1.10 & 89.1 & SQ limit \\
AZTSe      & 1.35 & 33.7 & 34.7 & 1.09 & 89.0 & SQ limit \\

\hline
CZTS       & 1.50 & 20.9 & 28.9 & 0.84 & 86.4 & TLC \\ 
CZTSe      & 1.00 & 20.3 & 47.7 & 0.53 & 81.0 & TLC \\
CZGSe      & 1.36 & 24.1 & 34.3 & 0.81 & 86.2 & TLC \\

CZTS:H     & 1.50 & 23.1 & 28.9 & 0.91 & 87.4 & TLC \\ 
CZTSe:H    & 1.00 & 23.7 & 47.7 & 0.60 & 82.7 & TLC \\
CZGSe:H    & 1.36 & 27.9 & 34.3 & 0.93 & 87.5 & TLC \\
AZTSe:H    & 1.35 & 30.8 & 34.7 & 1.01 & 88.1 & TLC \\
\hline

CZTS       & 1.50 & 11.0 & 21.7 & 0.73 & 69.27 & Exp.\cite{Yan:2018dw} \\
CZTSe      & 1.00  & 11.6  & 40.6   & 0.42  & 67.3  & Exp.\cite{Lee:2014cna} \\
CZTSSe     & 1.13 & 12.6 & 35.4 & 0.54 & 65.9  & Exp.\cite{Son:2019hm} \\
CZTGSe     & 1.11 & 12.3  & 32.3   & 0.53  & 72.7  & Exp.\cite{Kim:2016jr}  \\
CZGSe      & 1.36 & 7.6   & 22.8   & 0.56  & 60    & Exp.\cite{Choubrac:2018ex}  \\
AZTSe      & 1.35 & 5.2   & 21.0     & 0.50  & 48.7  & Exp.\cite{Gershon:2016kt}  \\ 
ACZCTS     & 1.40 & 10.1  & 23.4   & 0.65   & 66.2  & Exp.\cite{Hadke:2018hx} \\ 
    \end{tabular*}
    \caption{Device performance parameters of selected Cu and Ag kesterite solar cells and predicted by Shockley-Queisser limit and trap-limited conversion efficiency and found experimentally (Exp).
    }
    \label{tab:eff}
\end{table}

\begin{table}[ht]
\centering
\begin{tabular*}{\columnwidth}{lccccc}
        & $E_\mathrm{gap}$ & $\Delta V_\mathrm{OC}^\mathrm{nonrad}$ & $p_0$ & $\tau_\mathrm{SRH}$  & PLQY   \\ 
        & (eV)  & (V)      & (\si{cm^{-3}})  & (ns) & (\%)         \\ 
\hline
CZTS    & 1.50  & 0.39     & \num{3.3e15}    & 0.13 & \num{3.1e-05} \\ 
CZTSe   & 1.00  & 0.24     & \num{1.7e15}    & 3.4  & \num{9.8e-03} \\ 
CZGSe   & 1.36  & 0.29     & \num{9.0e15}    & 0.21 & \num{1.4e-03} \\ 
AZTSe   & 1.35  &          & \num{1.0e10}    &      &               \\ 
\hline
CZTS:H  & 1.50  & 0.32     & \num{3.8e16}    & 0.21 & \num{4.5e-04} \\ 
CZTSe:H & 1.00  & 0.17     & \num{1.8e16}    & 5.5  & \num{1.5e-01} \\ 
CZGSe:H & 1.36  & 0.17     & \num{4.8e17}    & 0.38 & \num{1.5e-01} \\ 
AZTSe:H & 1.35  & 0.08     & \num{1.7e16}    & 1130 & \num{4.6e-00} \\ 
\end{tabular*}
    \caption{Optoelectronic parameters derived from first principles of selected Cu and Ag kesterites. $\Delta V_\mathrm{OC}^\mathrm{nonrad}$ is the \VOC~ loss due to the nonradiative recombination, $p_0$ is the intrinsic hole concentration, $\tau _\mathrm{SRH}$ is the Shockley–Read–Hall lifetime and PLQY is the external photoluminescence quantum yield at 1-sun equivalent conditions.
    }
    \label{tab:par}
\end{table}

\section{Conclusions}

We have combined the physics of solar cells with modern first-principles defect theory
to assess the efficiency limit of solar cells.
We have included the thermal equilibrium concentrations of native defects of the absorber material, which reduces carrier lifetime,
and have proposed a first-principles method to calculate the maximum efficiency limited by recombination centers.
Sn-based kesterites suffer from severe nonradiative recombination due to native point defects.
The fast nonradiative recombination can be mitigated by
extrinsic doping and Ag-alloying, reducing the concentration of recombination centres,
thereby increasing the performance threshold to 29\%.

Although, our approach advances first-principles approaches for solar cells, its limitations should be noted.
We are pushing defect theory to its limits of applicability and note that inaccuracies, e.g. through finite-sized corrections or choice of exchange-correlation functional, will become magnified in the predictions of defect concentrations and capture cross-sections. 
The method inherits some of the limitations of the SQ approach.\cite{Guillemoles:2019gk}
It is based on bulk properties are therefore does not take into account surface or interface recombination.
Parasitic absorption effects in the buffer or window layers are also ignored. 

In the case of kesterite solar cells, although it is widely accepted that a short carrier life is the main performance bottleneck,\cite{Hages:2017co,Grenet:2018ia}
high series resistance can further reduce efficiency.\cite{Grenet:2018ia}
Thin-films are often inhomogeneous with lateral variations in stoichiometry.
Therefore, fluctuations of the band gap and the electrostatic potential can 
reduce the open-circuit voltage beyond our predictions.\cite{mattheis2008finite}

The TLC metric should be considered as an upper bound, based on the bulk properties of the absorber, that can be achieved when losses through other degradation pathways are minimal. 
In commercial photovolatic solar cells, \JSC~ and FF approach the SQ limit. The main efficiency-limiting factor is \VOC~ \cite{Nayak:2019im,Guillemoles:2019gk}, which we tackle.
Therefore, our method can provide a new direction for searching for promising photovoltaic materials
by providing a realistic upper limit on expected performance.
It can be used as part of screening procedures to select viable candidates.
Finally, we emphasise that to assess the genuine potential of real materials for photovoltaics, 
one should consider not only the thermodynamics of light and electrons, but also the thermodynamics of crystals.

\section{Data availability} 
The data that support the findings of this study are available in Zenodo repository with the identifier doi:10.5281/zenodo.XXXXXXX.
%
The code used in this study is available in Zenodo repository with the identifier doi:10.5281/zenodo.YYYYYYY.
We make use of the open-source packages \url{https://github.com/WMD-group/CarrierCapture.jl} (capture cross-sections), \url{https://github.com/jbuckeridge/sc-fermi} (equilibrium concentrations) and \url{https://github.com/jbuckeridge/cplap} (chemical fields).

\acknowledgements
We thank Samantha N. Hood, John Buckeridge, and Ji-Sang Park for valuable discussions.
This research has been funded by the EU Horizon2020 Framework (STARCELL, Grant No. 720907).
We are grateful to the UK Materials and Molecular Modelling Hub for computational resources, which is partially funded by EPSRC (EP/P020194/1).
Via our membership of the UK's HEC Materials Chemistry Consortium, which is funded by EPSRC (EP/L000202), this work used the ARCHER UK National Supercomputing Service (http://www.archer.ac.uk).
This work was also supported by a National Research Foundation of Korea (NRF) grant funded by the Korean government (MSIT) (No. 2018R1C1B6008728).

\bibliographystyle{rsc}
\bibliography{library}

\end{document}